# The effect of frames on engagement with quantum technology

Aletta L. Meinsma[1,2], Casper J. Albers[3], Pieter Vermaas[4], Ionica Smeets[1], Julia Cramer[1,2]

[1] Department of Science Communication & Society, Faculty of Science, Leiden University, Leiden, the Netherlands
[2] Leiden Institute of Physics, Faculty of Science, Leiden University, Leiden, the Netherlands
[3] Heymans Institute for Psychological Research, University of Groningen, Groningen, the Netherlands
[4] Philosophy Department, School of Technology, Policy & Management, Delft University of Technology, Delft, the Netherlands

**Abstract:** Quantum technology is predicted to have a significant impact on society once it matures. This study ($n$ = 637 adults representative of the Dutch population) examined the effect of different frames on engagement - specifically, information seeking, internal efficacy, general interest and perceived knowledge - with quantum technology. The different frames were: enigmatic, explaining quantum physics, benefit, risk and balanced. Results indicated that framing quantum as enigmatic does not affect engagement, while explaining quantum physics positively influences general interest. Furthermore, emphasising a benefit of quantum technology increases participants' internal efficacy, whereas highlighting both a benefit and a risk of quantum technology decreases perceived knowledge. Based on these findings, we offer practical advice for science communicators in the field and suggest further research.

## 1 Introduction

Quantum technology is an emerging technology which encompasses technologies like quantum computing, quantum networks and quantum sensors. All of these technologies are envisioned to have significant implications on society – both positively as well as negatively. Positive implications include, for instance, saving or extending lives through drug discovery (Busby et al., 2017; Outeiral et al., 2021), providing secure communication as a defence against hacking threats (Vermaas et al., 2019; Wehner et al., 2018), and identifying risks to ground conditions through accurate sensing (Stray et al., 2022). Examples of negative effects include concerns about how criminal organisations may use this technology (Busby et al., 2017; Vermaas et al., 2019), as well as unequal access to the technology depending on where you live (Holter et al., 2022). Holter et al. (2022) argue that simply allowing China, the West, and the wealthy to control quantum technology will likely worsen existing inequalities, and consequently deny populations access to the benefits that the technology can provide.

To maximise the positive and minimise the negative impacts of quantum technology on society, it is important to connect to society in an early phase of the technology's development (Roberson, 2021). Allowing for citizens to participate in societal dialogues could give technology developers and policy makers a better understanding of how quantum technology affects different groups of people (Roberson et al., 2021), and it could also lead to more public support and less public opposition (Kurath and Gisler, 2009; Mooney, 2010; Roberson et al., 2021). Moreover, citizens should be given the opportunity to participate from a democratic perspective, as this technology can significantly affect citizens' lives (Van Dam et al., 2020).

Engaging a larger, non-specialist audience with quantum technology poses challenges as the theory underlying quantum technology is very different to what we experience in our everyday lives. Societal actors could feel unqualified to participate in a dialogue about such a technology. Referring to quantum theory in certain ways, such as calling it enigmatic, potentially enhances this feeling even further (Coenen et al., 2022; Vermaas, 2017).

In this work, we discerned three different ways of communicating with non-experts about quantum technologies by means of *frames* used in the communication. These are: 1.) the enigmatic frame; 2.) the explanation frame; and 3.) the benefit, risk and balanced frames. The goal of our study was to determine the effect of these frames on the self-reported engagement of participants, as literature suggests that all of these affect engagement with quantum technology (Bobroff, 2017; Busby et al., 2017; Coenen et al., 2022; Retzbach and Maier, 2015; Vermaas, 2017). The next section covers the theoretical concepts on which our study is based in detail.

## 2 Theory

*2.1 Scientific engagement*

When we look at the history of science communication, there is a shift from focussing on public understanding of science to public engagement with science (Kurath and Gisler, 2009). According to Kurath & Gisler's (2009) historic overview, governments tried to increase public acceptance of science and technology by educating the public through one-way communication models (i.e., referred to as the deficit model) starting from the 1980s. However, this approach did not always succeed in increasing public acceptance. Some groups became even more critical after obtaining additional information from experts, for example in the United Kingdom on genetically modified crops (Van Dam et al., 2020).

Science communication scholars have since advocated moving to public engagement models (also called "upstream" engagement; Kurath and Gisler, 2009; Mooney, 2010; Rogers-Hayden and Pidgeon, 2007; Van der Sanden and Meijman, 2008), but this remains challenging. The domain of science and technology is alienating for some people (Brooks, 2017; Druckman and Bolsen, 2011), which hinders engagement. For example, Brooks (2017) found through a qualitative thematic analysis on students' reflective writings that a lot of the students thought science was inaccessible. Also, many of them did not identify as scientists even while some of them had a scientific job themselves. Research into factors that could increase engagement with science therefore remains important.

Public engagement with science can be measured in a variety of ways, but in our study, we have decided to adopt the scientific engagement measure outlined by Shulman et al. (2020). They found that avoiding jargon has a positive indirect effect on participant's self-reported scientific engagement, which they measured via four variables. They assessed participants' intentions to seek additional information on the scientific technologies presented (self-driving cars, medical printers and robot surgeons), their beliefs about their own ability to understand and engage with information about the scientific technologies, their general interest in them, and their confidence in their knowledge of the scientific technologies. In this way, different dimensions of scientific engagement were examined.

*2.2 The effect of frames on engagement*

In this work, we examined the effect of *frames* on quantum technology engagement. Framing refers to "select[ing] some aspects of a perceived reality and mak[ing] them more salient in a communicating context, in such a way to promote a particular problem definition, causal interpretation, moral evaluation, and/or treatment recommendation" (Entman, 1993: 52). By emphasising only some aspects while ignoring other parts of certain information, frames can influence how audiences interpret and perceive that information (Entman, 1993).

We studied multiple frames occurring around quantum technology, which previous (theoretical) research suggested could affect engagement with quantum technology. These are: 1.) the enigmatic frame (Coenen et al., 2022; Vermaas, 2017); 2.) the explanation frame (Bobroff, 2017); and 3.) the benefit, risk and balanced frames (Busby et al., 2017; Retzbach and Maier, 2015).

*2.2.1 The enigmatic frame*

The first frame we studied is the enigmatic frame. According to Coenen et al. (2022) and Vermaas (2017), the enigmatic frame could make the domain of quantum technology unapproachable for non-specialist audiences. If quantum experts highlight that they find quantum science and technology enigmatic, non-experts might consequently think that there is no chance that they will ever be able to acquire the knowledge needed to participate in a dialogue about the technology. According to Meinsma et al. (2023), almost a third of the quantum experts in their dataset (30%) framed quantum as something spooky or enigmatic in their popularised science talk. The frame is therefore clearly visible in popular communication about quantum science and technology, which leads to the question if it indeed forms a barrier for engagement.

*2.2.2 The explanation frame*

Secondly, we examined the effect of the explanation frame on engagement with quantum technology. The three quantum physics concepts that are often explained by quantum experts (Meinsma et al., 2023) and journalists (Meinsma et al., 2024) in their communications towards a broader audience are *superposition*, *entanglement* and *contextuality*. *Superposition* refers to the fact that a particle following the laws of quantum mechanics can be in two states simultaneously (i.e., a particle in a superposition state is in a linear combination of states; Griffiths, 2014; Nielsen & Chuang, 2010). *Entanglement* means that particles share a state, making it no longer appropriate to refer to those particles as separate (Griffiths, 2014; Nielsen and Chuang, 2010). Finally, *contextuality* addresses the idea that performing a measurement on a quantum state has an irreversible impact on that state (although there is more nuance to this, see Jaeger, 2019).

To our knowledge, no previous study has examined the effect of explaining quantum physics concepts on engagement. On the one hand, the explanation frame could improve engagement. Working on over a hundred outreach projects about quantum and solid-state physics, Bobroff (2017) noticed that people are fascinated by the theory and not only by potential applications. On the other hand, the counter intuitiveness of these concepts may also make people feel alienated by the field. Performing a measurement on a quantum state, for instance, does not appear in everyday experiences: for us, looking at an object does not all of a sudden change its properties, but by 'looking' at a quantum object you do change its properties. This is "genuine quantum weirdness", according to Van Wezel et al. (2023). Therefore, in a similar way as the enigmatic frame (Coenen et al., 2022; Vermaas, 2017), this frame could hinder engagement. The question thus arises whether the frame encourages engagement or acts as a barrier to it.

*2.2.3 The benefit, risk and balanced frames*

Thirdly, we investigated the benefit, risk and balanced frames on participants' engagement with quantum technology. The benefit, risk and balanced frames focus on discussing one of the benefits of quantum technology, one of its risks, or one of its benefits and risks at the same time. By providing the benefit, risk and balanced frames, the personal relevance of the technology might become clear which in turn could increase engagement.

According to several studies, e.g., Druckman & Bolsen (2011), people will typically have little knowledge about emergent technologies and lack incentives to become informed, because the personal relevance of learning more will be unclear at best. Several non-specialist audiences indicated that they view the personal relevance for learning more about quantum technology to be low (Busby et al., 2017; Moraga-Calderón et al., 2020). Moraga-Calderón et al. (2020) found that high school students expressed quantum technology to be important for society, but not for themselves. The researchers rediscovered the expression "*Important, but not for me*".

A public dialogue exercise in 2017 in the United Kingdom revealed in a more qualitative way that talking about the benefits of quantum technology increased participants' engagement and excitement about the technology (Busby et al., 2017). The authors observed that, at first, the participants who entered the dialogue found quantum science and technology difficult to understand and for experts only. The personal relevance of learning about quantum technology was not immediately clear. However, when talking about benefits of the technology, participants became more engaged and excited about quantum technology as they started to understand how quantum technology could be relevant to their own lives. The benefit that quantum technology might save or extend lives and the risk of misusing quantum technology for cyberwarfare and hacking got the most reactions.

Simultaneously presenting both a benefit and a risk, known as a balanced frame, is a way of conveying scientific uncertainty (Retzbach and Maier, 2015). The way in which scientific uncertainty is presented determines whether it has positive or negative effects in science communication (Gustafson and Rice, 2020). When presenting scientific uncertainty by highlighting both the potential benefits as well as the potential harms, it was found to lead to slight positive effects: Retzbach & Maier (2015) found that in the case of nanotechnology, individuals open to considering new or inconsistent information became more interested in new technology. This illustrates that the balanced frame can positively influence public engagement with new technology via an increase in interest.

To our knowledge, a more quantitative assessment of the impact of the benefit, risk and balanced frames on various dimensions of quantum technology engagement is absent. Our goal is therefore to investigate the effect of these frames on quantum technology engagement. We expect these frames to positively affect engagement because of the personal relevance becoming clear.

*2.3 Research questions*

As introducing public engagement in an early phase of a technology's development is important (Kurath and Gisler, 2009; Mooney, 2010; Priest, 2010), it is essential to investigate the incentives and barriers to engagement with quantum technology early on. Our goal is to examine the effect of different frames on engagement with quantum technology. Our study aims to answer the following research questions (RQs):

RQ1. Does the enigmatic frame influence the self-reported engagement with quantum technology?
RQ2. Does the explanation frame influence the self-reported engagement with quantum technology?
RQ3. Does the benefit, risk and balanced frames influence the self-reported engagement with quantum technology?

## 3 Method

Following our research questions, we used a survey design in which participants were randomly assigned in a 2 (enigmatic, not enigmatic) x 2 (explanation, no explanation) x 4 (none, benefit, risk, balanced) between-subjects design. A total of *n* = 649 participants were recruited by KiesKompas (https://www.kieskompas.nl/en/), a Dutch survey company, to form a representative sample of Dutch adults in age, gender and education. Before the data collection, the study information, design plan, sampling plan, variables and analysis plan were preregistered on OSF Registries (https://osf.io/h6b7s). The Ethics Review Committee of the Faculty of Science, Leiden University gave ethical approval to conduct the study (reference number 2023 – 02).

*3.1 Independent variables*

The independent variables consisted of the enigmatic frame condition, the explanation frame condition and the benefit, risk and balanced frames conditions. The specific wordings per experimental condition can be found in Table A1 in the Appendix. Jargon terms such as superposition, entanglement and contextuality were avoided (as jargon has been found to hinder engagement; Shulman et al., 2020).

*3.1.1   Enigmatic frame condition*

In the enigmatic frame condition, participants viewed a statement in which quantum mechanics was referred to as enigmatic ("raadselachtig" in Dutch), while in the no enigmatic frame condition this adjective was left out.

*3.1.2   Explanation frame condition*

To obtain a more rigorous test of whether providing a quantum physics explanation affects engagement, we opted for a message sampling approach (Slater et al., 2015). This approach means that, rather than a single message, various messages that all share the trait being studied are used as experimental stimuli to limit the effect of one specific message. In this way, we gained confidence in that we measured the effect of providing an explanation of a quantum physics concept on participants' engagement, rather than the effect of a certain concept or a certain type of explanation. We designed six different messages that explained a quantum physics concept: participants were randomly assigned to read an explanation about superposition, entanglement or contextuality in two different ways. Afterwards, we collapsed the results.

*3.1.3   Benefit, risk and balanced frames condition*

In addition to both manipulations above, participants were also randomly assigned to view a benefit frame, a risk frame, both a benefit and risk frame (i.e., a balanced frame), or none of these frames. We designed a benefit frame around health and a risk frame around security (as these got the most reactions in the public dialogue exercise; Busby et al., 2017).

*3.2 Dependent variable*

After reading the text associated with one of the 16 experimental conditions, participants were asked to self-report their engagement with quantum technology. We adapted the 4-variable engagement measure of Shulman et al. (2020) such that all of the statements referred to quantum technology instead of science and technology in general. The statements per variable can be found in Table A2 in the Appendix.

Our quantum technology engagement measure consisted of 16 statements which participants were asked to answer to on a 5-point Likert scale (ranging from strongly disagree – strongly agree). The first variable assessed whether participants intended to seek out more information about quantum technology in the future (information seeking, a 3-item scale, Cronbach's alpha = 0.94). The second measure assessed whether participants' believed that they can understand and engage with information about quantum technology (internal efficacy, a 4-item scale, Cronbach's alpha = 0.859). The third measure was to assess participants' interest toward quantum technology (general interest, a 6-item scale, Cronbach's alpha = 0.892). And fourthly, we were interested in the confidence that the participants have about their own quantum technology knowledge (perceived knowledge. This scale originally had 4 items, with Cronbach's alpha = 0.446. A leave-one-out-analysis (cf. Nunnally, 1978) showed that the 3 item scale with items 1, 2 and 4, performed considerably better, Cronbach's alpha = 0.848, thus we work with these three items).

*3.3 Analysis and statistical procedures*

Data were analysed with R (version 4.2.2, RStudio Team, 2022). To evaluate the effects of the different conditions, we conducted a linear multiple regression analysis with dummy variables. We tested for main effects and did not analyse possible interactions. The model was run for each engagement variable: information seeking, internal efficacy, general interest and perceived knowledge. The sum score for each of the engagement variables was calculated by adding up the Likert scale questions associated with them. A condition was determined to be a significant contributor to the particular engagement variable if *p* < 0.0125 (based on an alpha level of 0.05 and a Bonferroni correction for multiple testing). We also conducted a MANOVA to test the effect of the conditions on a linear combination of information seeking, internal efficacy, general interest and perceived knowledge.

# 4 Results

*4.1 Descriptives*

The survey ran between May 16th and May 23rd 2023 and resulted in a total sample of *n* = 649 Dutch-speaking adults. The median time to complete the survey was 3 minutes and 18 seconds. As 12 participants finished the survey in 90 seconds or less, we assumed that they did not think their response through and we discarded them from the analysis. This resulted in a final sample of *n* = 637 participants (394 men, 243 women; $M_{age}$ = 57.53 years, sd = 15.98 years).

To approximate the population of the Netherlands, the data was weighted by post-stratification for gender, age and education. The weights were trimmed at 98th percentile, to lessen the effect of the highest-weight outliers. Table 1 shows the number of participants that were assigned to each condition in the 2 (enigmatic, not enigmatic) x 2 (underlying explanation, no underlying explanation) x 4 (none, benefit, risk, balanced) design for the unweighted and weighted data.

**Table 1**
Number of participants in each condition before and after weighting.

|  | Unweighted | | Weighted | |
| --- | --- | --- | --- | --- |
|  | Did not read frame | Read frame | Did not read frame | Read frame |
| **Enigmatic frame** | 321 | 316 | 308.655 | 323.328 |
| **Explanation frame** | 319 | 318 | 318.180 | 313.804 |
| **Risk frame** | 309 | 328 | 294.669 | 337.314 |
| **Benefit frame** | 319 | 318 | 324.853 | 307.131 |

*4.2 Significant contributors to the regression models*

The results of the regression models for information seeking, internal efficacy, general interest and perceived knowledge are shown in Table 2. First of all, the results show that the enigmatic frame condition did not significantly contribute to any of the four models.

Secondly, the explanation condition was found to be a significant contributor for general interest ($b = 1.171$, $p = 0.004$). Participants that viewed a quantum physics explanation scored 1.171 points higher (on the five point scale) on general interest than the group of participants that had not viewed a quantum physics explanation. The explanation condition did not contribute significantly to the other models for information seeking, internal efficacy and perceived knowledge.

Thirdly, the benefit frame was found to be a significant contributor to the model for internal efficacy ($b = 1.060$, $p = 0.011$), while the risk frame did not significantly contribute to any of the four models. Furthermore, the score of the participants that received a balanced frame significantly dropped in comparison to participants that only read a risk or only read a benefit frame for perceived knowledge ($b = -1.040$, $p = 0.011$). None of the other engagement variables were significantly affected by the risk and/or benefit frames.

**Table 2**
Multiple linear regression analysis of the experimental conditions per outcome variable. *p*-Values in bold indicate significance at a level of <0.0125.

| | | | | Information seeking |
|---|---|---|---|---|
| | Estimate | Std. error | *t*-value | *p*-value |
| Intercept | 7.624 | 0.320 | 23.793 | **<0.001** |
| Enigmatic frame | -0.412 | 0.251 | -1.640 | 0.101 |
| Explanation frame | 0.258 | 0.251 | 1.027 | 0.305 |
| Risk frame | 0.310 | 0.352 | 0.880 | 0.379 |
| Benefit frame | 0.699 | 0.368 | 1.899 | 0.058 |
| Risk frame × Benefit frame | -0.850 | 0.504 | -1.687 | 0.092 |

| | | | | Internal efficacy |
|---|---|---|---|---|
| | Estimate | Std. error | *t*-value | *p*-value |
| Intercept | 8.527 | 0.362 | 23.581 | **<0.001** |
| Enigmatic frame | -0.527 | 0.283 | -1.858 | 0.064 |
| Explanation frame | 0.310 | 0.284 | 1.091 | 0.276 |
| Risk frame | 0.542 | 0.397 | 1.366 | 0.173 |
| Benefit frame | 1.060 | 0.415 | 2.552 | **0.011** |
| Risk frame × Benefit frame | -1.392 | 0.569 | -2.449 | 0.015 |

| | | | | General interest |
|---|---|---|---|---|
| | Estimate | Std. error | *t*-value | *p*-value |
| Intercept | 17.669 | 0.522 | 33.867 | **<0.001** |
| Enigmatic frame | -0.774 | 0.409 | -1.892 | 0.059 |
| Explanation frame | 1.171 | 0.409 | 2.861 | **0.004** |
| Risk frame | 0.626 | 0.573 | 1.093 | 0.275 |
| Benefit frame | 0.916 | 0.599 | 1.529 | 0.127 |
| Risk frame × Benefit frame | -1.132 | 0.820 | -1.380 | 0.168 |

| | | | | Perceived knowledge |
|---|---|---|---|---|
| | Estimate | Std. error | *t*-value | *p*-value |
| Intercept | 6.003 | 0.259 | 23.142 | **<0.001** |
| Enigmatic frame | -0.142 | 0.203 | -0.698 | 0.486 |
| Explanation frame | -0.042 | 0.203 | -0.206 | 0.837 |
| Risk frame | 0.344 | 0.285 | 1.208 | 0.227 |
| Benefit frame | 0.673 | 0.298 | 2.259 | 0.024 |
| Risk frame × Benefit frame | -1.040 | 0.408 | -2.549 | **0.011** |

*Note.* Intercept is the expected sum score for the given variable for when none of the experimental conditions apply.

*4.3 MANOVA*

Finally, we tested whether the conditions correlated significantly with a linear combination of information seeking, internal efficacy, general interest and perceived knowledge. As shown in Table 3, the explanation condition correlated significantly (Pillai's trace = 0.028, $F$(4, 626) = 4.491, $p$ = 0.001) whereas the other conditions did not.

**Table 3**
MANOVA model to test the effect of the conditions on a linear combination of the four engagement variables: information seeking, internal efficacy, general interest and perceived knowledge. *p*-Values in bold indicate significance at a level of <0.05.

|  | Pillai | Approx. F | Num Df | Den Df | *p*-value |
|---|---|---|---|---|---|
| Enigmatic frame | 0.003 | 0.437 | 4 | 626 | 0.782 |
| Explanation | 0.003 | 0.477 | 4 | 626 | 0.753 |
| Risk frame | 0.010 | 1.622 | 4 | 626 | 0.167 |
| Benefit frame | 0.028 | 4.491 | 4 | 626 | **0.001** |
| Risk frame × Benefit frame | 0.012 | 1.825 | 4 | 626 | 0.122 |

# 5 Discussion

This study examined the effect of the enigmatic frame, the explanation frame, and the benefit, risk and balanced frames on the self-reported engagement with quantum technology. To the best of our knowledge, this is the first study that experimentally assessed the effect of these frames on quantum technology engagement.

*5.1 No effect of the enigmatic frame on engagement*

First of all, we found that the enigmatic frame did not significantly affect any of the dimensions of engagement we studied. This is surprising, given that Coenen et al. (2022) and Vermaas (2017) argued that the frame would have negative implications for people's beliefs about their own ability to understand and engage with quantum technology. Apparently the frame is not as detrimental as expected for people's engagement with quantum technology. Our finding is encouraging, given the regular occurrence of the enigmatic frame in popular communication about quantum science and technology (Meinsma et al., 2023, 2024).

It should be noted that participants in the enigmatic frame condition scored lower on all four dimensions of engagement than participants not in this condition, but these results are not significant. As participants in the enigmatic frame condition were exposed to the frame only once, this raises questions about the possible effects of repeated exposure of the frame.

*5.2 The explanation frame affects general interest*

Secondly, we found that an explanation of a quantum physics concept increased participants' engagement and especially interest in quantum technology. Our finding is in line with the observation from Bobroff (2017) who noticed that people are fascinated by quantum physics concepts. We conclude from this that the counter intuitiveness of these concepts do not have negative effects, but positively influence quantum technology engagement. Given that many experts (Meinsma et al., 2023) and journalists (Meinsma et al., 2024) explain at least one of the three concepts we included in our study - superposition, entanglement, and contextuality - in their popular communications, our finding is encouraging.

*5.3 The benefit frame increases internal efficacy, whereas the balanced frame decreases perceived knowledge*

Thirdly, the benefit frame had a positive effect on participants' internal efficacy. This result supports the conclusion from the public dialogue exercise in the United Kingdom that discussing benefits raises participants' engagement and excitement about quantum technology as the personal relevance becomes clear (Busby et al., 2017). To further facilitate societal engagement with quantum technology, research should look into other factors that allow the personal relevance of quantum technology to become clear.

We furthermore found that none of the engagement variables were significantly affected by the risk frame. This is similar to the findings in Retzbach & Maier's (2015) study, where the risk frame did not significantly affect public engagement with nanotechnology. Apparently, the risk frame does not urge people to seek further information on the topic, nor does it affect any of the other engagement dimensions.

Finally, considering the balanced frame, participants' perceived knowledge decreased significantly compared to participants exposed to either a benefit or a risk frame. This raises concerns about presenting a balanced perspective on quantum technology to public engagement, which contradicts the conclusions of Retzbach & Maier (2015). Retzbach & Maier (2015) found that the balanced frame had no detrimental effects on public engagement with nanotechnology, and even found slight positive effects for participants open to considering new or inconsistent information. In our study, we did not look at effects from differences in personality traits, which may be a reason for this discrepancy. It would be interesting to explore the effects that factors other than frames, such as different personality traits, age, gender and science credibility, have on quantum technology engagement.

*5.4 Limitations of this study*

The lab setting of this study allowed us to accurately control the experiment. However, a translation of our experiment into a real-world setting would give better insights into real-life effects. Future studies could involve writing different news stories about quantum technology, each emphasizing a specific frame, and then tracking how readers respond to these articles. Another research idea is to develop different quantum technology outreach activities and test whether there are differences in participants' engagement in the activity. This way we could get a better understanding of how our controlled experiment compares to a real-world setting.

Our study used Shulman et al.'s (2020) scientific engagement measure, which includes the four variables: information seeking, internal efficacy, general interest and perceived knowledge. However, it can be argued that public engagement with science includes more elements, such as beliefs about science and trust in scientists (see Retzbach & Maier, 2015). Additionally, there are limitations to self-reporting, as participants may give socially desirable answers or assess themselves inaccurately. Therefore, future research should also measure actual behavioural engagement. For example, by including an option to receive additional information about quantum technology by clicking on a link, thereby measuring people's actual information-seeking behaviour.

Finally, our sample involved Dutch-speaking adults recruited by KiesKompas. This demographic scope could limit the generalizability of our findings to other cultural or linguistic contexts. Therefore, we advise future research to replicate our study in other countries.

*5.5 Practical implications*

In conclusion, our results give support to the following: communications that are effective in increasing public engagement with quantum technology explain quantum physics concepts, emphasise the benefits, but avoid presenting both the benefits and risks. This conclusion leads to an interesting tension, as there have been pleas for media coverage to provide sufficient attention on both the benefits as well as the risks of quantum technology (see for example Roberson et al., 2021). Widely reflecting on both the benefits and risks of quantum technology is important to ensure quantum technology will mostly bring public benefit (Roberson et al., 2021). In our opinion, this ethical consideration outweighs the fact that a balanced frame leads to a decrease in the perceived knowledge dimension of engagement. Therefore, we would still advise science communicators to present both the benefits as well as the risks of quantum technology.

# Funding

We acknowledge funding from the Dutch Research Council (NWO) through a Spinoza Grant awarded to R. Hanson (Project Number SPI 63-264) and thank Ronald Hanson for this opportunity. This work was supported by the Dutch National Growth Fund (NGF), as part of the Quantum Delta NL programme.

# Acknowledgements

We thank KiesKompas for the pleasant cooperation. Moreover, we are grateful to Hillary Shulman for the informative email conversation about their scientific engagement measure. Finally, we thank Sanne Romp for having a thorough read through the manuscript.

# Appendix

**Table A1**
Wordings for the 16 experimental conditions

| Nr | Group | Wordings for Experimental Conditions (in Dutch) | Wordings for Experimental Conditions (translated to English) |
|---|---|---|---|
| 1 | Control group | Control group statement: "Quantumtechnologie is een opkomende technologie. Quantumtechnologie gebruikt wetenschappelijke kennis die de allerkleinste deeltjes beschrijft, zoals elektronen." | Control group statement: "Quantum technology is an emerging technology. Quantum technology uses scientific knowledge that describes the smallest particles, such as electrons." |
| 2 | Enigmatic frame group | Enigmatic frame statement: "Quantumtechnologie is een opkomende technologie. Quantumtechnologie gebruikt wetenschappelijke kennis die de allerkleinste deeltjes, zoals elektronen, op een raadselachtige manier beschrijft." | Enigmatic frame statement: "Quantum technology is an emerging technology. Quantum technology uses scientific knowledge that describes the smallest particles, such as electrons, in a mysterious way." |
| 3 | Explanation of a quantum concept group<br>1a) Superposition - definition<br>1b) Superposition - analogy<br>2a) Entanglement - definition<br>2b) Entanglement - analogy<br>3a) Contextuality - definition<br>3b) Contextuality - analogy | Control group statement +<br><br>One of the following 6 statements (Explanation frame statement):<br><br>1a) "Zo zegt deze wetenschap dat het voor kleine deeltjes mogelijk is om op twee plaatsen tegelijkertijd te zijn. Dit is iets dat we in ons dagelijks leven niet zien. In ons dagelijks leven kan iets maar op één plek tegelijkertijd zijn."<br><br>1b) "Zo zegt deze wetenschap dat het voor kleine deeltjes mogelijk is om op twee plaatsen tegelijkertijd te zijn. Stel dat dit ook voor een auto zou gelden. Dan zou een auto op twee verschillende wegen tegelijk kunnen | Control group statement +<br><br>One of the following 6 statements (Explanation frame statement):<br><br>1a) "For example, this science says that it is possible for small particles to be in two places at the same time. This is something that we don't see in our daily lives. In our daily lives, something can only be in one place at the time."<br><br>1b) "For example, this science says that it is possible for small particles to be in two places at the same time. Suppose this also applies to a car. Then a car could drive on two different roads at the same time, for example on |

rijden, bijvoorbeeld op de A2 en tegelijkertijd op de A28. Dit is iets dat we in ons dagelijks leven niet zien. In ons dagelijks leven rijdt een auto óf op de A2 óf op de A28."

2a) "Zo zegt deze wetenschap dat twee kleine deeltjes die ver weg zijn, toch sterk met elkaar verbonden zijn. Als de toestand van het ene deeltje verandert, verandert de toestand van het andere deeltje onmiddellijk mee. Dit is iets wat we in ons dagelijks leven niet zien."

2b) "Zo zegt deze wetenschap dat twee kleine deeltjes in een gezamenlijke toestand kunnen zijn. Daardoor heeft het geen zin meer om over deze twee deeltjes te praten alsof het twee afzonderlijke deeltjes zijn. Stel dat dit ook voor twee gekleurde ballen zou gelden. Als de ballen in een gezamenlijke toestand zijn en je kijkt naar de kleur van één bal, dan kan dat onmiddellijk de kleur van de andere bal beïnvloeden. Ook al is die andere bal mijlenver weg."

3a) "Zo zegt deze wetenschap dat door naar een deeltje te kijken, de eigenschappen van dat deeltje kunnen veranderen. Dit is iets wat we in ons dagelijks leven niet ervaren. In ons dagelijkse leven verandert iets niet van eigenschap, zoals van kleur of vorm, alleen maar door ernaar te kijken."

3b) "Zo zegt deze wetenschap dat door naar een deeltje te kijken, de eigenschappen van dat deeltje kunnen veranderen. Stel dat dit ook voor een trui zou gelden: zodra u naar de trui kijkt is deze bijvoorbeeld groen. Als u wegkijkt, en een stuk later opnieuw naar de trui kijkt, the A2 and at the same time on the A28. This is something we don't see in our daily lives. In our daily lives, a car either drives on the A2 or the A28."

2a) "For example, this science says that two small particles that are far away are still strongly connected. If the state of one particle changes, the state of the other particle changes immediately. This is something we don't see in our daily lives."

2b) "For example, this science says that two small particles can be in a joint state. As a result, it no longer makes sense to talk about these two particles as if they were two separate particles. Suppose this also applies to two coloured balls. If the balls are in a joint state and you look at the colour of one ball, it can immediately affect the colour of the other ball. Even though the other ball is miles away."

3a) "For example, this science says that by looking at a particle, the properties of that particle can change. This is something we do not experience in our daily lives. In our daily lives, something does not change its properties, such as colour or shape, just by looking at it."

3b) "For example, this science says that by looking at a particle, the properties of that particle can change. Suppose this also applies to a sweater: as soon as you look at the sweater it is for example green. If you look away and look at the sweater again a little later, the

| | | | |
|---|---|---|---|
| | | is de trui opeens grijs. Het kijken naar de trui heeft een fysiek effect op die trui." | sweater is suddenly grey. Looking at the sweater has a physical effect on that sweater." |
| 4 | Benefit frame group | Control group statement + | Control group statement + |
| | | Benefit frame statement: "Sommige wetenschappers zeggen dat quantumtechnologie in de toekomst levens kan gaan redden of verlengen. Volledig ontwikkelde quantumtechnologie heeft namelijk de potentie om nieuwe medicijnen te ontwerpen." | Benefit frame statement: "Some scientists say that quantum technology could save or extend lives in the future. Fully developed quantum technology has the potential to design new medicines." |
| 5 | Risk frame group | Control group statement + | Control group statement + |
| | | Risk frame statement: "Sommige wetenschappers zeggen dat quantumtechnologie in de toekomst veiligheidsproblemen kan gaan veroorzaken. Volledig ontwikkelde quantumtechnologie heeft namelijk de potentie om gebruikt te worden voor cyberoorlogsvoering." | Risk frame statement: "Some scientists say that quantum technology could cause safety problems in the future. Fully developed quantum technology has the potential to be used for cyber warfare." |
| 6 | Balanced frame group | Control group statement + | Control group statement + |
| | | Balanced frame statement: "Sommige wetenschappers zeggen dat quantumtechnologie in de toekomst levens kan gaan redden of verlengen. Volledig ontwikkelde quantumtechnologie heeft namelijk de potentie om nieuwe medicijnen te ontwerpen. Andere wetenschappers zeggen dat quantumtechnologie in de toekomst veiligheidsproblemen kan gaan veroorzaken. Volledig ontwikkelde quantumtechnologie heeft namelijk de potentie om gebruikt te worden voor cyberoorlogsvoering." | Balanced frame statement: "Some scientists say that quantum technology could save or extend lives in the future. Fully developed quantum technology has the potential to design new medicines. Other scientists say that quantum technology could cause safety problems in the future. Fully developed quantum technology has the potential to be used for cyber warfare." |
| 7 | Enigmatic frame + Explanation of a quantum concept group | Enigmatic frame statement + Explanation of a quantum concept statement | Enigmatic frame statement + Explanation of a quantum concept statement |
| 8 | Enigmatic frame + Benefit frame group | Enigmatic frame statement + Benefit frame statement | Enigmatic frame statement + Benefit frame statement |
| 9 | Enigmatic frame + Risk frame group | Enigmatic frame statement + Risk frame statement | Enigmatic frame statement + Risk frame statement |
| 10 | Enigmatic frame + Balanced frame group | Enigmatic frame statement + Balanced frame statement | Enigmatic frame statement + Balanced frame statement |
| 11 | Explanation of a quantum concept + Benefit frame group | Control group statement + Explanation of a quantum concept statement + Benefit frame statement | Control group statement + Explanation of a quantum concept statement + Benefit frame statement |
| 12 | Explanation of a quantum concept + Risk frame group | Control group statement + Explanation of a quantum concept statement + Risk frame statement | Control group statement + Explanation of a quantum concept statement + Risk frame statement |
| 13 | Explanation of a quantum concept + Balanced frame group | Control group statement + Explanation of a quantum concept statement + Balanced frame statement | Control group statement + Explanation of a quantum concept statement + Balanced frame statement |
| 14 | Enigmatic frame + Explanation of a quantum concept + Benefit frame group | Enigmatic frame statement + Explanation of a quantum concept statement + Benefit frame statement | Enigmatic frame statement + Explanation of a quantum concept statement + Benefit frame statement |
| 15 | Enigmatic frame + Explanation of a quantum concept + Risk frame group | Enigmatic frame statement + Explanation of a quantum concept statement + Risk frame statement | Enigmatic frame statement + Explanation of a quantum concept statement + Risk frame statement |

| 16 | Enigmatic frame + Explanation of a quantum concept + Balanced frame group | Enigmatic frame statement + Explanation of a quantum concept statement + Balanced frame statement | Enigmatic frame statement + Explanation of a quantum concept statement + Balanced frame statement |

**Table A2**
Engagement questions

| Outcome variable | Cronbach's alpha | Statements (in Dutch) | Statements (translated to English) |
|---|---|---|---|
| Information seeking | 3-item scale Alpha = 0.94 | 1. Ik ben van plan om binnenkort informatie over quantumtechnologie op te zoeken | 1. I plan to seek information about quantum technology in the near future |
| | | 2. Ik zal proberen om in de komende tijd informatie over quantumtechnologie op te zoeken | 2. I will try to seek information about quantum technology in the near future |
| | | 3. Het is mijn bedoeling om meer te weten te komen over quantumtechnologie | 3. I intend to find out more information about quantum technology |
| Internal efficacy | 4-item scale Alpha = 0.859 | 1. Ik denk dat ik goed in staat ben om deel te nemen aan discussies over quantumtechnologie | 1. I consider myself to be well qualified to participate in discussions about quantum technology |
| | | 2. Ik heb het gevoel dat ik een redelijk goed begrip heb van de belangrijke kwesties rond quantumtechnologie waarmee Nederland wordt geconfronteerd | 2. I feel that I have a pretty good understanding of the important quantum technology issues facing the country |
| | | 3. Ik heb het gevoel dat ik net zo goed een oordeel kan leveren op het gebied van quantumtechnologie als de meeste andere mensen | 3. I feel that I could do as good a job in the quantum technology field as most other people |
| | | 4. Ik denk dat ik beter geïnformeerd ben over quantumtechnologie dan de meeste mensen | 4. I think that I am better informed about quantum technology than most people |
| General interest | 6-item scale Alpha = 0.892 | 1. Ik wil meer leren over quantumtechnologie | 1. I am interested in learning about quantum technology |
| | | 2. Ik vind het debat rond quantumtechnologie interessant | 2. I find the debate surrounding quantum technology interesting |

| | | 3. Ik wil mij gaan verdiepen in quantumtechnologie | 3. I want to learn more about quantum technology |
| | | 4. Ik ben nieuwsgierig geworden naar quantumtechnologie | 4. Quantum technology is exciting |
| | | 5. Ik vind quantumtechnologie saai (reverse-coded) | 5. I find quantum technology boring (reverse-coded) |
| | | 6. Wat ik net heb gelezen over quantumtechnologie levert stof tot nadenken | 6. These quantum technological ideas were thought-provoking |
| Perceived knowledge | 4-item scale We worked with items 1, 2 and 4 resulting in alpha = 0.848 | 1. Ik heb kennis over quantumtechnologie | 1. I am knowledgeable about quantum technology |
| | | 2. Ik voel me goed geïnformeerd over zaken rond quantumtechnologie | 2. I am well-informed about issues related to quantum technology |
| | | 3. Ik weet niet zoveel als ik zou willen weten over zaken rond quantumtechnologie (reverse-coded) | 3. I don't know as much as I'd like to know about the issues surrounding quantum technology (reverse-coded) |
| | | 4. Ik heb vertrouwen in mijn begrip rond quantumtechnologie | 4. I trust my knowledge about quantum technology |